\documentclass[letterpaper, 10 pt, conference]{ieeeconf}  

\IEEEoverridecommandlockouts                              

\usepackage{graphicx}
\usepackage{multirow}

\usepackage{graphicx}
\usepackage{xcolor}
\usepackage{fancyhdr}

\title{\LARGE \bf
Multi-Phase Cross-modal Learning for Noninvasive Gene Mutation Prediction in Hepatocellular Carcinoma
}

\author{
Jiapan Gu$^{*1}$, Ziyuan Zhao$^{*1}$, Zeng Zeng$^{\dagger 1}$, Yuzhe Wang$^{2}$, Zhengyiren Qiu$^{2}$, Bharadwaj Veeravalli$^{2}$\\
Brian Kim Poh Goh$^{3}$, Glenn Kunnath Bonney$^{4}$, Krishnakumar Madhavan$^{4}$\\
Chan Wan Ying$^{5}$, Lim Kheng Choon$^{6}$, Thng Choon Hua$^{5}$, and Pierce KH Chow$^{7}$
\thanks{* The first two authors contribute equally in this work. $^{\dagger}$ Corresponding author. The work was supported by Singapore-China NRF-NSFC Grant (Grant No. NRF2016NRF-NSFC001-111) and Pre-GAP Grant, Singapore (Grant No. ACCL/19-GAP023-R20H) with special acknowledgement to all collaborators in the NMRC Translational and Clinical Research (TCR) Flagship Programme (NMRC/TCR/015-NCC/2016). $^{1}$ Institute for Infocomm Research (I2R), Agency for Science, Technology and Research (A*STAR), Singapore. $^{2}$ National University of Singapore, Singapore. $^{3}$ Singapore General Hospital, Department of Hepatopancreatobiliary and Transplant Surgery, Singapore. $^{4}$ National University of Hospital, Division of Hepatobiliary and Pancreatic Surgery, Singapore. $^{5}$ National Cancer Centre, Division of Oncologic Imaging, Singapore. $^{6}$ Singapore General Hospital, Department of Vascular and Interventional Radiology, Singapore. $^{7}$ National Cancer Centre, Division of Surgery and Surgical Oncology, Singapore.}}

\begin{document}

\maketitle
\thispagestyle{empty}
\pagestyle{empty}

\thispagestyle{fancy}
\fancyhead{}
\lfoot{}
\lfoot{\scriptsize{Copyright 2020 IEEE. Published in the 2020 42nd Annual International Conference of the IEEE Engineering in Medicine and Biology Society (EMBC), scheduled for July 20-24, 2020 at the Montréal, Canada. Personal use of this material is permitted. However, permission to reprint/republish this material for advertising or promotional purposes or for creating new collective works for resale or redistribution to servers or lists, or to reuse any copyrighted component of this work in other works, must be obtained from the IEEE. Contact: Manager, Copyrights and Permissions / IEEE Service Center / 445 Hoes Lane / P.O. Box 1331 / Piscataway, NJ 08855-1331, USA. Telephone: + Intl. 908-562-3966.}}
\rfoot{}

\begin{abstract}
Hepatocellular carcinoma (HCC) is the most common type of primary liver cancer and the fourth most common cause of cancer-related death worldwide. Understanding the underlying gene mutations in HCC provides great prognostic value for treatment planning and targeted therapy. Radiogenomics has revealed an association between non-invasive imaging features and molecular genomics. However, imaging feature identification is laborious and error-prone. In this paper, we propose an end-to-end deep learning framework for mutation prediction in APOB, COL11A1 and ATRX genes using multiphasic CT scans. Considering intra-tumour heterogeneity (ITH) in HCC, multi-region sampling technology is implemented to generate the dataset for experiments. Experimental results demonstrate the effectiveness of the proposed model.
\newline

\indent \textit{Clinical relevance}— The proposed framework applied deep learning and multi-region sampling to predict gene mutations based on multi-phase 3D CT scans, achieving 75\% average accuracy. The architecture can be implemented to provide clinical diagnostic and predictive oncology services.
\end{abstract}


\section{INTRODUCTION}

Liver cancer is one of the leading causes of cancer-related deaths in many parts of the world and one of the most common cancers among males in Singapore~\cite{sgweb2020}. Hepatocellular carcinoma (HCC) is the most common type of primary liver cancer which is the sixth most common cancer in the world and the fourth leading cause of cancer mortality globally~\cite{yang2019global}. HCC is known to be a highly heterogeneous disease~\cite{dragani2010risk}, especially in genetic level, which means the lack of consistency in the therapeutic outcome and thus may lead to the difficulty in clinic for designing targeted therapy and precision medicine. Recent studies have identified several mutations in genes associated with HCC. For instance, telomerase reverse transcriptase (TERT) promoter, TP53/p53 and CTNNB1/beta-catenin have been identified as the most commonly mutated genes in HCC~\cite{rao2016frequently}.

\begin{figure}[t]
    \centering
    \includegraphics[width = 7.7 cm]{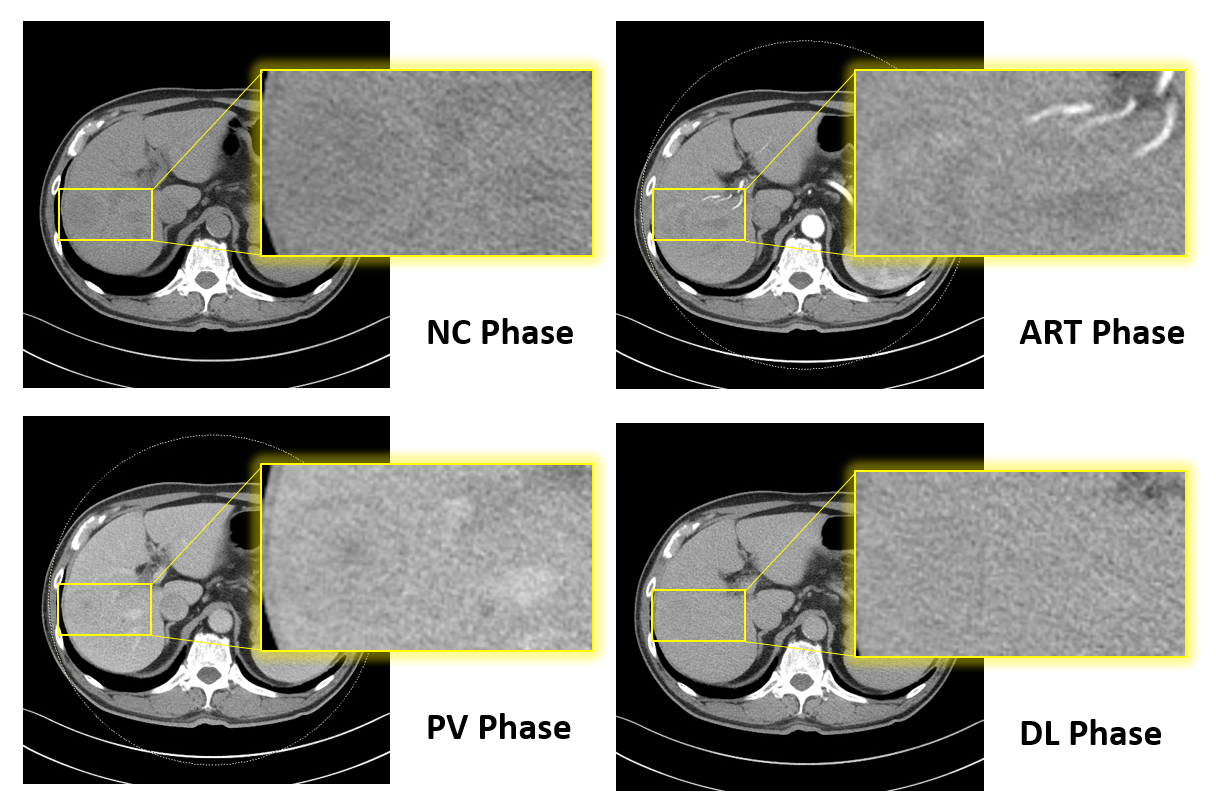}
    \caption{Illustration of the appearance of tumors in different phases. The appearance of the tumor and non-tumor (background) in NC and PV phase is similar. However, the appearance of the tumor and non-tumor is quite different in ART phase.}
    \label{fig:multiphase}
\end{figure}

In general, current methods of genome sequencing require liver biopsy and specialized equipment, which limit the routine usage. Besides, these invasive sampling methods are destructive, and can not provide multiple continuous snapshots. Many research work have demonstrated that different imaging features in computed tomography (CT) scans are correlated with the global gene expression of primary human liver cancer, {\textit{e}.\textit{g}.}, Segal~\textit{et al.} used 28 features out of over 100 image features manually extracted from CT scans to reconstruct $78\%$ of the gene expression patterns in HCC patients~\cite{segal2007decoding}. Such an integrated diagnostic approach between Radiology and Genomics is termed as ``radiogenomics"~\cite{rutman2009radiogenomics}.

However, the development of radiogenomics for HCC is critically impeded by the following challenges. First, Dynamic CT images are widely used in liver CT scanning. Multi-phase CT scans are divided into four phases, {\textit{i}.\textit{e}.}, non-contrast (NC) phase, arterial (ART) phase, portal venous (PV) phase, and delay (DL) phase in Fig.~\ref{fig:multiphase}. Varying object sizes, {\textit{e}.\textit{g}.}, lesions and tumors make it difficult to obtain comprehensive information in dynamic multi-modal images. Second, there is intra-tumour heterogeneity (ITH) in HCC~\cite{zhai2017spatial}, which means gene mutations are different in different parts of tumors. This scenario makes machine learning approaches, especially convolutional neural network (CNN) very difficult to train for gene mutation prediction in HCC, as the input slices of the whole tumor with different gene mutations are assigned the same mutation label. Third, radiogenomics requires quantitative imaging features annotated by expert radiologists, which is laborious, time-consuming and suffers from high intra / inter-observer variance. 

In this work, we propose a deep CNN framework to address the challenges of automatic gene mutation prediction in HCC. More specifically, a multi-stream CNN is applied for multi-phase cross-modal feature extraction followed by an aggregation layer to effectively fuse and utilize 4D information. Besides, some image traits (biomarkers)~\cite{ocker2018biomarkers} identified by radiologists are aggregated as auxiliary information for decision making. Extensive experiments are implemented on the dataset collected from different hospitals in Singapore, in which, multi-region sampling is applied to avoid mismatch problems in ITH. Experimental results and analysis demonstrate the effectiveness of the proposed framework for mutation prediction in APOB, COL11A1, and ATRX genes, which is easily implemented to predict more gene mutations if sufficient data are provided.


\section{RELATED WORK}
\subsection{Radiogenomics and AI}
As a rapidly developing field, radiogenomics has shown potential value for diagnostic and therapeutic strategies. In the fields of radiomics and radiogenomics, high-throughput extraction of qualitative and quantitative imaging features from radiographs is required to obtain diagnostic, predictive, or prognostic information~\cite{kumar2012radiomics}. To explore the relationships between gene expression and imaging, Segal~\textit{et al.}~\cite{segal2007decoding} defined `units of distinctiveness', termed as `traits' from qualitative imaging features of liver cancer and reconstructed nearly 80\% of the global gene expression profiles using 28 image traits. Aerts~\textit{et al.}~\cite{aerts2014decoding} proposed a quantitative strategy for the correlation between CT images and genome data, which reflects great clinical significance. These contribute to the development of emerging technologies such as computer vision and deep learning in radiogenomics~\cite{li2018novel, smedley2018using}. The use of artificial intelligence (AI) for genomics and molecular profiling of cancers will be hugely beneficial as it is non-invasive and captures a comprehensive view of the tumor. However, most of the current radiogenomics analysis in liver cancer extracted image features from single-phase images which did not consider the changes of shapes and sizes of tumors across phases. Besides, merging multiphasic information remains a critical issue.

\subsection{Multi-region Sampling}

\begin{figure}[tb]
    \centering
    \includegraphics[width = 7.5 cm]{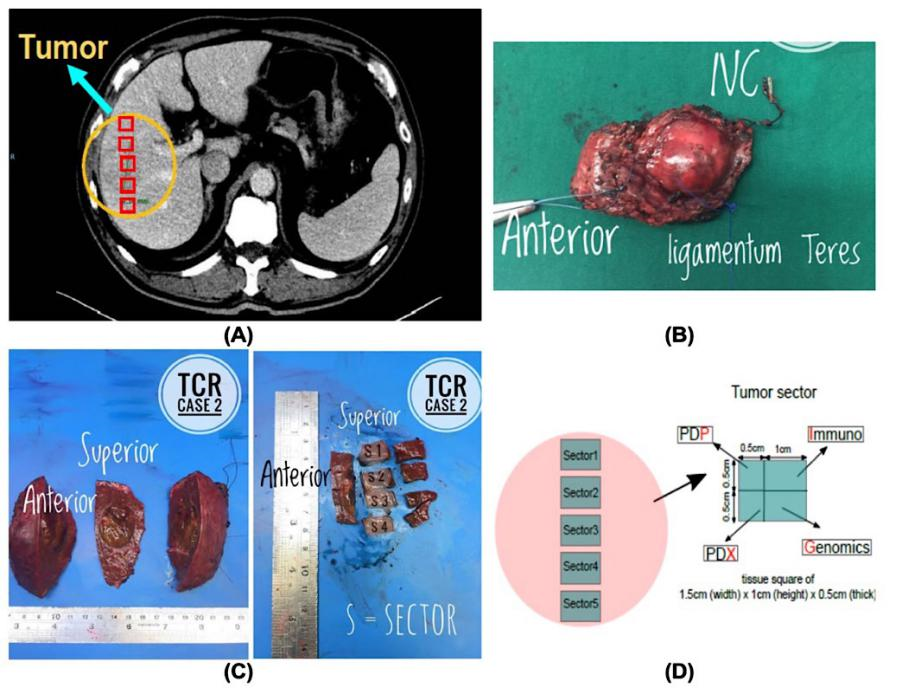}
    \caption{Multi-region sampling for genomics. (A) Tumor and sectors are annotated on the CT scan. (B) Resected tumor with annotated orientation. (C) Multi-region sampling of tumor over the central slice with annotated sectors. (D) Illustration of multi-region sampling and obtaining samples for the omics levels of analysis in PLANET~\protect\footnotemark[1]. PDP and PDX refer to patient-derived progenitor cells and patient-derived xenografts respectively.}
    \label{fig:multiregion}
\end{figure}

Genomic profiling methods behind most of the current radiogenomics analysis rely on single biopsy samples, which often reflects a part of the tumor. These methods significantly underestimate intratumoral genomic heterogeneity in cancers especially in HCC. A large proportion of HCC displays a clear geographic segregation where spatially closer sectors are genetically more similar~\cite{zhai2017spatial}. Furthermore, this genetic heterogeneity in HCC influences the training process of CNNs. Slices with different mutations are put into networks with the same mutation label, which misleads the training process of networks. Zhai~\textit{et al.}~\cite{zhai2017spatial} first carried out research on intra-tumour heterogeneity (ITH) in HCC using multi-region sampling. As shown in Fig.~\ref{fig:multiregion}, first, tumor and sectors were annotated on the CT scans. Then a central slice is cut from the patient tumor, and a linear grid of tumor sectors is then harvested and examined for further multiomics analysis. The dataset used in this work adopted multi-region sampling.

\footnotetext[1]{More details of multi-omics discovery platform PLANET are shown in: https://www.nccs.com.sg/research-innovation/research-programmes/disease-focus-liver}

\begin{figure*}[ht]
    \centering
    \includegraphics[width=14.2 cm]{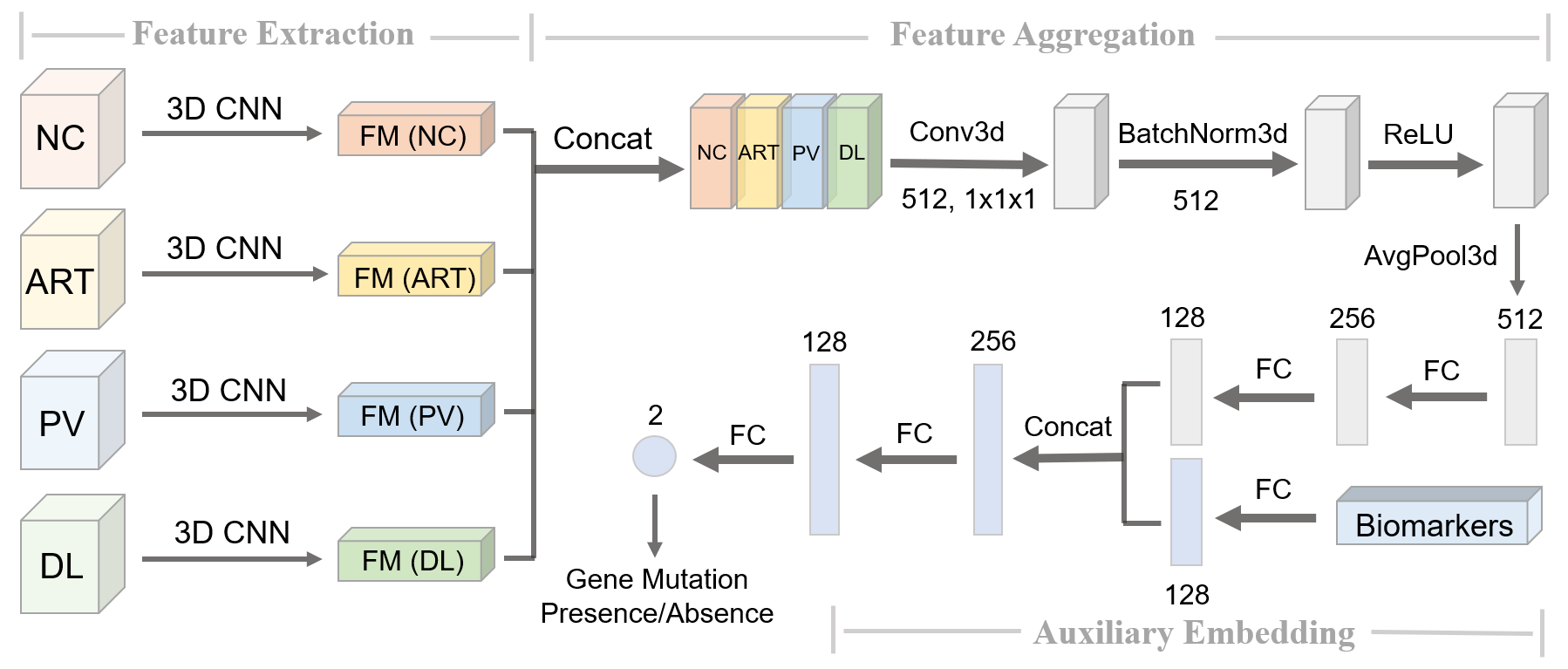}
    \caption{Overall framework of the proposed method, the processed images from different phases are trained using four separate 3D CNN (3D ResNet-18), followed by concatenation and a shadow CNN block. The biomarkers are embedded with image features for final prediction on gene mutation.}
    \label{fig:archi}
\end{figure*}


\section{METHODOLOGY}
The proposed method for gene mutation prediction is illustrated in Fig.~\ref{fig:archi}, which consists of three main components: (i) Multi-stream Feature Extraction, (ii) Feature Aggregation, and (iii) Auxiliary Embedding. First, the processed images from different phases are put into four deep CNN blocks separately for feature extraction; then a lightweight CNN is applied to fuse multi-stream information for further prediction~\cite{DLCase}. For better prediction, biomarkers annotated by doctors are embedded into the network as auxiliary information to improve the prediction accuracy.

In multi-stream feature extraction, \textit{Residual Neural Network} (ResNet)~\cite{resnet,zhao2019bira} is implemented for deep feature extraction, in which some layers are skipped through shortcut connection to avoid the vanishing gradient problem. The 3D ResNet-18 pre-trained on ImageNet~\cite{deng2009imagenet} is applied.

In feature aggregation, feature maps from four streams are combined by concatenation along the channel dimension, followed by a shallow network including one convolution layer ($512, 1\times1\times1$), one batch normalization layer, and one ReLU activation layer for dimensionality reduction in the channel dimension space.

Compared to using images only, some image traits (biomarkers) are embedded into the final fully connected layer. These image traits are the characteristics described by radiologists, which can be regarded as indicators of physiologic and pathologic processes in response to various diagnostic or therapeutic procedures. In our experiments, nine kinds of biomarkers are annotated as binary variables by doctors, such as ``presence of intra-tumoral vessels". More details can be found in Fig.~\ref{fig:correlation}. These biomarkers are treated as a series of binary numbers followed by a fully connected layer with a size of 128. The output of the fully connected layer is combined with image features for further prediction. Binary cross-entropy is used as the loss function for the binary classification problem.



\section{EXPERIMENTS}
\subsection{Dataset and implementation}

\begin{figure}[thb]
    \centering
    \includegraphics[width =7.5 cm]{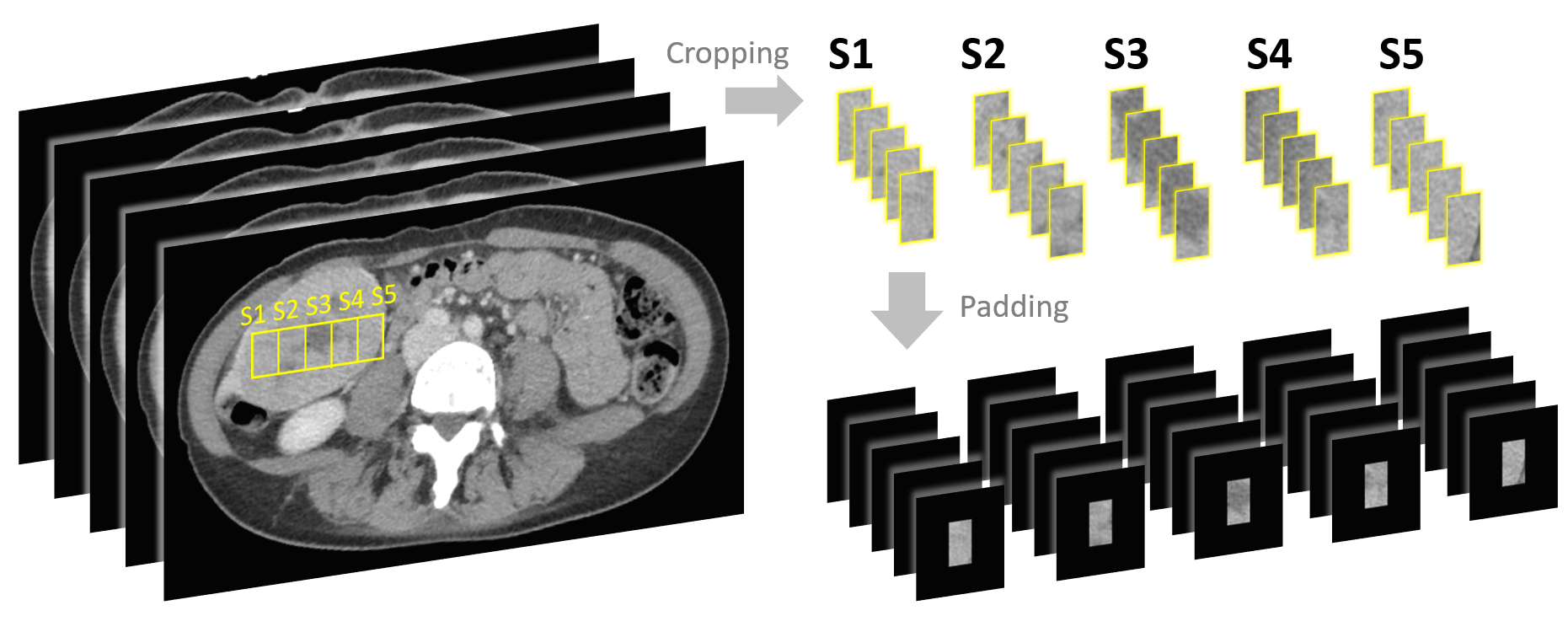}
    \caption{Image preprocessing pipeline: first, five continuous slices of the same sectors are extracted, and then, the regions of the sectors are cropped from the original slices followed by padding with zeros, keeping the same size in length and width.}
    \label{fig:Preprocessing}
\end{figure}

The dataset has been collected from multiple hospitals located in Singapore, which consists of 3D multiphasic CT scans, genomics information and biomarker sequences from $27$ patients with the approval of the Institutional Review Board. Due to various sources of data, $14$ patients with all four phases were considered in our experiments. The DNA sequencing was based on multi-region sampling, in which, gene mutations are different from different sectors of the tumor, therefore, our training samples are on the basis of the sector rather than patient. The ground truth mutation labels were extracted from the DNA sequencing, and the corresponding sectors were extracted from the patients' CT scans. Because of various sizes of sectors, the sectors were padded with zeros to keep the same size in length and width, and five adjacent slices along $z$-axis based on the central slice of sector were cropped from volumetric images to probe the spatial information along the third dimension, see Fig.~\ref{fig:Preprocessing}. Typically, HCC in different regions of the liver is clinically scored on the basis of their CT image features, such as size and margin. Therefore, nine biomarkers were annotated by radiologists based on CT scans.

Because of data imbalance in different gene mutations, prevalent mutated genes, {\textit{i}.\textit{e}.}, APOB, COL11A1 and ATRX are considered in this study.  Given that the mutations are not mutually exclusive, a separate model per mutation is used to alleviate the problem of data imbalance among mutations. We compare our model with single-stream CNN using single-phase CT scans, in which, ResNet-18 is trained on images from one of the four phases, {\textit{e}.\textit{g}.}, Single Stream (ART phase). To explore the effectiveness of auxiliary embedding in the proposed architecture, the proposed model without biomarkers is also implemented, {\textit{i}.\textit{e}.}, Proposed (without biomarkers). To validate the stability of the proposed model, leave-one-out cross validation (LOOCV) is performed, in which, each patient is excluded from the training set one at a time and then classified on the basis of the predictor built from the data for all the other patients.




\begin{table}[thb]
\caption{Automatic gene mutation prediction performance of different methods.}
\label{results}
\begin{center}
\begin{tabular}{c|c|c|c}
\hline
\multicolumn{1}{c|}{\multirow{2}{*}{Methods}} & \multicolumn{3}{|c}{Accuracy} \\ \cline{2-4} 
\multicolumn{1}{c|}{}                        & APOB    & COL11A1    & ATRX   \\ \hline
Single Stream (NC phase)                            & 58.9\%    & 51.3\%    & 60.1\%   \\ 
Single Stream (ART phase)                           & 70.1\%    & 55.5\%    & 57.5\%   \\ 
Single Stream (PV phase)                            & 64.1\%    & 56.4\%    & 68.1\%   \\ 
Single Stream (DL phase)                            & 68.4\%    & 56.9\%    & 64.5\%   \\ \hline
Proposed (without biomarkers)                                     & 77.3\%    & 67.7\%    & 73.4\%   \\ 
Proposed                   & 76.7\%    & 71.7\%    & 76.5\%   \\ \hline
\end{tabular}
\end{center}
\end{table}
\subsection{Results and discussion}

Table~\ref{results} summarizes the results of all methods on the dataset. The proposed method outperforms all other methods in different gene mutations, which demonstrates the effectiveness of feature fusion cross phases. Besides, biomarkers can help to improve the accuracy of mutation prediction in COL11A1 and ATRX, which proves the feasibility of biomarkers in radiogenomics analysis.

\begin{figure}[thb]
    \centering
    \includegraphics[width = 7.8cm]{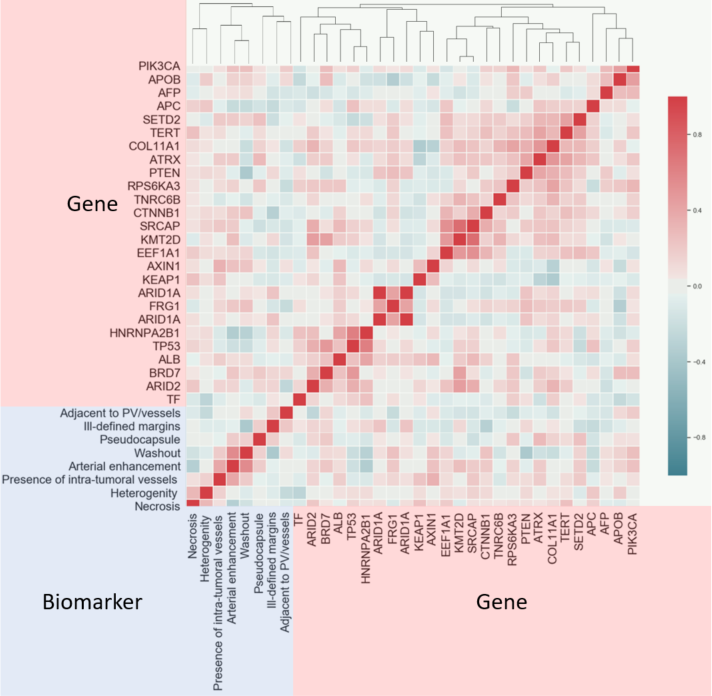}
    \caption{Correlation map and hierarchical clustering based on pairwise Pearson correlation coefficients of biomarkers and gene mutations. Some gene mutations are correlated (red) or anti-correlated (blue).}
    \label{fig:correlation}
\end{figure}

To further explore the correlations among biomarkers and gene mutations, a correlation map is generated with the whole dataset, as shown in Fig.~\ref{fig:correlation}. We find that most of these biomarkers are correlated with gene mutations. Furthermore, some gene mutations are correlated with each other. We performed hierarchical clustering of the correlation map and found that some gene mutations are grouped into the same cluster, therefore, the complex relationships among gene mutations can be modeled using graph neural networks for prediction in the future~\cite{chen2019gated}.

\section{CONCLUSIONS AND FUTURE WORK}

In this work, we discuss the challenges and standard-of-care imaging technologies in HCC radiogenomics analysis. Considering the intra-tumour heterogeneity (ITH) in HCC, we propose a sector-based multi-stream cross-modal deep learning framework for mutation prediction in genes. Multiphasic CT scans are processed and extracted by multi-stream CNN followed by feature aggregation. Moreover, the biomarkers are embedded into the final layer for further prediction. Experimental results on the dataset show the effectiveness of the proposed framework on mutation prediction in APOB, COL11A1, and ATRX genes. However, our framework is extendable to more gene mutations with sufficient training data. The correlation between the gene mutations and biomarkers not only validate the predictive value of biomarkers, but also show the significant correlations among different gene mutations. In our future work, the relationships between gene mutations will be analyzed and considered as predictor variables.



\bibliographystyle{IEEEbib}
\bibliography{refs.bib}

\end{document}